\documentclass[reprint,amsmath,nofootinbib,amssymb,aps,pra]{revtex4-2}
\usepackage{amsmath}%
\usepackage{amsfonts}%
\usepackage{amssymb}%
\usepackage[cal=cm]{mathalfa}
\usepackage{graphicx}
\usepackage{dcolumn}
\usepackage{bm}
\usepackage{braket}
\usepackage{color}
\usepackage{bbm}
\usepackage{nicefrac}
\usepackage{float}
\usepackage{placeins}
\usepackage[dvipsnames]{xcolor}
\usepackage[utf8]{inputenc}
\usepackage[normalem]{ulem}
\usepackage{nicefrac}
\usepackage{dsfont}
\usepackage{xcolor, soul} 
\usepackage[normalem]{ulem} 
\usepackage{siunitx}
\usepackage[citebordercolor=blue]{hyperref}
\usepackage[utf8]{inputenc}
\usepackage{gensymb}
\usepackage{mathtools}
\usepackage[rightcaption]{sidecap}

\begin{document}

\title{Spin to Orbital Angular Momentum Conversion in the Neutron Nucleus Weak Interaction}

\author{Niels Geerits}
\email{niels.geerits@geerits.at}
\affiliation{%
Atominstitut, Technische Universit\"at Wien, Stadionallee 2, 1020 Vienna, Austria}
\author{Stephan Sponar}
\affiliation{%
Atominstitut, Technische Universit\"at Wien, Stadionallee 2, 1020 Vienna, Austria}
\author{Hartmut Lemmel}
\affiliation{%
Atominstitut, Technische Universit\"at Wien, Stadionallee 2, 1020 Vienna, Austria}
\affiliation{Institut Laue-Langevin, 71 Avenue des Martyrs, CS 20156, 38042
Grenoble Cedex 9, France}
\author{Andrea Piovano}
\affiliation{Institut Laue-Langevin, 71 Avenue des Martyrs, CS 20156, 38042
Grenoble Cedex 9, France}
\author{Kyle E Steffen}
\affiliation{Centre for Exploration of Energy and Matter, Indiana University, Bloomington, Indiana 47408, USA}
\author{William Michael Snow}
\affiliation{Centre for Exploration of Energy and Matter, Indiana University, Bloomington, Indiana 47408, USA}

\date{\today}
\hyphenpenalty=800\relax
\exhyphenpenalty=800\relax
\sloppy
\setlength{\parindent}{0pt}
\begin{abstract} 
The neutron nucleus weak interaction generates a small parity non conserving rotation of the neutron spin as the neutron moves through matter. This parity-odd neutron rotary power can reach up to $ 2.2\cdot10^{-4} \ \mathrm{rad\cdot cm}^{-1}$ for slow neutrons in $^{139}$La, which possesses a p-wave resonance at 0.7 eV where the parity-odd asymmetry is greatly amplified. This spin rotation must conserve total angular momentum. Here we show that this conservation occurs through neutron orbital angular momentum generation. We experimentally demonstrate that a neutron prepared initially in a longitudinal spin-orbit state experiences no change in total angular momentum.
\end{abstract}

\maketitle

\section{Introduction.} 
Orbital angular momentum (OAM) states of free particles were first realized in photons~\cite{Allen1992,Enk1994,Terriza2007} and later also in electrons~\cite{McMorran2011,Guzzinati2014,Grillo2014}. In 2015 extrinsic orbital angular momentum was demonstrated in neutrons as well~\cite{Clark2015}. Here extrinsic implies that the OAM is collective beam property as opposed to each individual neutron carrying some quantum OAM around its propagation axis. From there it was proposed to generate spin textured beams consisting of a lattice of spin vortices~\cite{Sarenac2018}, using magnetic prisms. This method was later generalized to twisted neutron phase textures in interferometry~\cite{Geerits2023}. Finally, in 2022 intrinsic neutron OAM was demonstrated using arrays of silicon fork gratings~\cite{sarenac2022experimental}. \par 
In photons~\cite{Marrucci2006} and neutrons~\cite{Sarenac2018,Sarenac2019} the spin angular momentum degree of freedom can be entangled with the OAM degree of freedom. We refer to these as spin-orbit states. In all of these cases the quantization axis of the orbital angular momentum is parallel to the neutron propagation direction (longitudinal). In a recent article we proposed that the Aharonov-Casher~\cite{Aharonov1984}/Schwinger term~\cite{Schwinger1948} can induce not only longitudinal, but also transverse spin orbit states where the OAM axis is perpendicular to the propagation direction~\cite{Geerits2021}. These beams are of scientific interest for new types of scattering experiments that they can enable~\cite{Afanasev2019}. Beams carrying spin and OAM can be sensitive to special types of interesting material properties such as chiral structures~\cite{Brullot2016}.
\par In this paper we show that the parity-odd neutron-nucleus weak interaction should induce longitudinal and transverse spin-orbit entangled states. The weak interaction terms in the Hamiltonian do not conserve orbital and spin angular momentum separately, so by angular momentum conservation any change in spin angular momentum must be compensated by a change in OAM. By passing a spin-orbit textured neutron beam through a $^{139}\mathrm{La}$ sample, we show that the weak interaction conserves the total angular momentum of the beam. Note that here the OAM is carried by the beam, analogous to~\cite{Clark2015,Sarenac2019}. \par 
Previous experiments have observed parity-odd neutron spin rotation due to the weak interaction~\cite{Forte1980,Heckel1984,Heil1999}. To our knowledge, neutron OAM in this system has not yet been investigated.
We describe a method to demonstrate experimentally that parity violation in neutron-nucleus transmission generates a small but nonzero orbital angular momentum and demonstrate this phenomenon using the 0.7 eV p-wave resonance of the $^{139}$La nucleus, one of the greatest amplifiers of neutron-nucleus parity violation ever measured~\cite{Heckel1984,Heil1999}. Such a beam will possesses spin-orbit entanglement. This phenomenon can enable the future generation of new types of single-particle entangled neutron beams with up to 4 different quantum variables: neutron spin, neutron orbital angular momentum, neutron energy, and neutron position. Such a multiply-entangled state of a single neutron has to our knowledge been created or investigated experimentally. The generation of a 4 degree of freedom entangled neutron state is within reach experimentally if we can add orbital angular momentum generators to the existing technologies employed in neutron interferometry~\cite{Rauch1974} or neutron spin echo interferometric techniques such as Spin Echo Small Angle Neutron Scattering (SESANS)~\cite{Rekveldt1996,Gaehler1996,Rekveldt2005}. Triple entanglement has already been observed and used for tests of quantum contextuality in both types of instruments \cite{Hasegawa2010,Shen2020}.\par 
A positive observation of parity-odd spin-orbit generation would imply that the theory of polarized neutron optics in the presence of parity violation for paraxial rays will need to be revised as no such terms are present in the existing formulations. Such a revision may have important implications for planned experiments to search for P-odd and T-odd interactions in neutron-nucleus interactions using forward transmission. One suggested neutron nucleus interaction which violates P and T symmetry is a coupling between neutron spin and nuclear polarization with the form $\vec{\sigma}\cdot[\vec{p}\times\vec{I}]$, with $\vec{\sigma}$ the pauli spin matrices, $\vec{p}$ the neutron momentum and $\vec{I}$ the nuclear polarization. Note the similarity to the Schwinger term~\cite{Schwinger1948} where theory has already calculated nonzero OAM generation, which suggests that the OAM carried by the neutron may in fact contribute to the scattering cross section.
\section{Theoretical Framework}
Neutron interactions in the medium which depend on the neutron spin $\vec{\sigma}$ exhibit phenomena directly analogous to optical activity and circular dichroism. Shortly after the parity-violating $V-A$ structure for the weak interaction current was established, F. C. Michel~\cite{Michel1964} suggested that the neutron forward scattering amplitude $f(0)$ in matter could possess a parity-odd component of the form $f_{pnc}(0)=G \vec{\sigma} \cdot \vec{p}$ where $\vec{\sigma}$ is the neutron spin, $\vec{p}$ is the neutron momentum, and  $G$ is on the order of the Fermi constant $G_{F}$. In turn this interaction can produce parity-odd contribution to neutron optical activity and neutron dichroism of the form $d\phi_{pnc}/dz=-4\pi N Re(G)$ and $A_{z}=(\sigma_{+}-\sigma_{-})/(\sigma_{+}+\sigma_{-})=4\pi Im(G)/\sigma$ where $d\phi_{pnc}/dz$ is the parity-odd neutron optical rotary power and $A_{z}$ is the parity-odd neutron longitudinal asymmetry. These expressions are model-independent results that come from nonrelativistic scattering theory~\cite{Stodolsky}.

Parity violation in neutron optics from neutron-nucleus interactions has been developed and described in several works~\cite{Zaretsky1983, Flambaum1984, Noguera1986, Flambaum1992, Flambaum1993DynamicalEO}. In the nonrelativistic limit one can write the effective Hamiltonian for P-odd neutron interactions with a nucleus in a form first proposed by Michel~\cite{Michel1964} and employed in later work~\cite{Karl1977}  

\begin{equation}
H=p^{2}/2m +V(r)+{1 \over 2} \{ F(r), {\vec{\sigma} \cdot \vec{p}}  \} 
\end{equation}

where $\sigma$ is the neutron spin, $\vec{p}$ is the neutron momentum, and $F(r)$ is the weak potential. The weak interaction Hamiltonian is given by the anti-commutator between the weak potential, $F(r)$, and the neutron helicity operator $\vec{\sigma}\cdot\vec{p}$ 

\begin{equation}\label{WeakHam}
    H_{weak}=\{F(r),\vec{\sigma}\cdot\vec{p}\}=F(r)\vec{\sigma}\cdot\vec{p}-iF_r(r)\vec{\sigma}\cdot\hat{r}
\end{equation}

where the subscript $r$ indicates the first derivative with respect to r. The parity-odd operator $\vec{\sigma}\cdot\vec{p} $ in the first term is the one which leads to both the helicity dependence of the total cross section $P$ and the parity-odd neutron rotary power $d\phi/dz$ when the matrix element of the Hamiltonian between plane wave states is taken to calculate the forward scattering amplitude \cite{Karl1977, Forte1978}. 


The large parity-odd asymmetry $A_{L}=0.1$  seen in the 0.7 eV p-wave resonance in $^{139}$La, which is about 6 orders of magnitude larger than one would expect based on the strength of the NN weak interaction,  is understood to come from standard model parity violation effects as amplified near p-wave resonances in the dense forest of narrow neutron-nucleus resonances nin heavy nuclei. 
This amplification mechanism on p-wave neutron-nucleus resonances comes from the mixing of s-wave and p-wave compound nucleus resonances and was well enough understood to be predicted theoretically~\cite{Sushkov1980, Sushkov1982} before it was confirmed by experiment~\cite{Forte1980, Alfimenkov1983, Alfimenkov1984, Heckel1984} and later measured with higher precision at the 0.7 eV p-wave resonance in $^{139}$La  \cite{Yuan1991, Heil1999, Haseyama2002}. The Time Reversal Invariance and Parity at Low Energies (TRIPLE) collaboration~\cite{Mitchell1999zz, Mitchell2001} measured the parity-odd helicity dependence of the total cross section for 75 different resonances at greater than $3\sigma$ statistical significance in several stable nuclei with atomic mass numbers between $93<A<130$ and near $A=230$. After applying a statistical analysis of this broad set of data using ideas from nuclear chaos, they concluded that all of these observed effects were consistent with standard model parity violation.

\subsection{OAM Generation in Cold and Thermal Neutrons}
When scattering from a bulk sample we can average $F(r)$ over the spatial coordinate, which produces an effective bulk potential with the form $V=b\vec{\sigma}\cdot\vec{p}$. The gradient of $F(r)$ naturally vanishes after this averaging. Thus a neutron traversing through such a bulk sample can be written as
\begin{equation}\label{Hamiltonian}
	H=-\frac{\hbar^2}{2m}\nabla^2 + b\vec{\sigma}\cdot\vec{p}
\end{equation}
 In the limit of low neutron energies far away from neutron-nucleus resonances this approximation can be more rigorously justified through an argument given by Stodolsky~\cite{stodolsky1974neutron}. Note that in general $b$ is complex with the real part contributing to parity non conserving spin rotation and the imaginary part to spin dichroism. The latter is much weaker~\cite{kolomensky1981observation} and thus will be neglected in this work. 
\begin{figure}
\includegraphics[width=\textwidth/2]{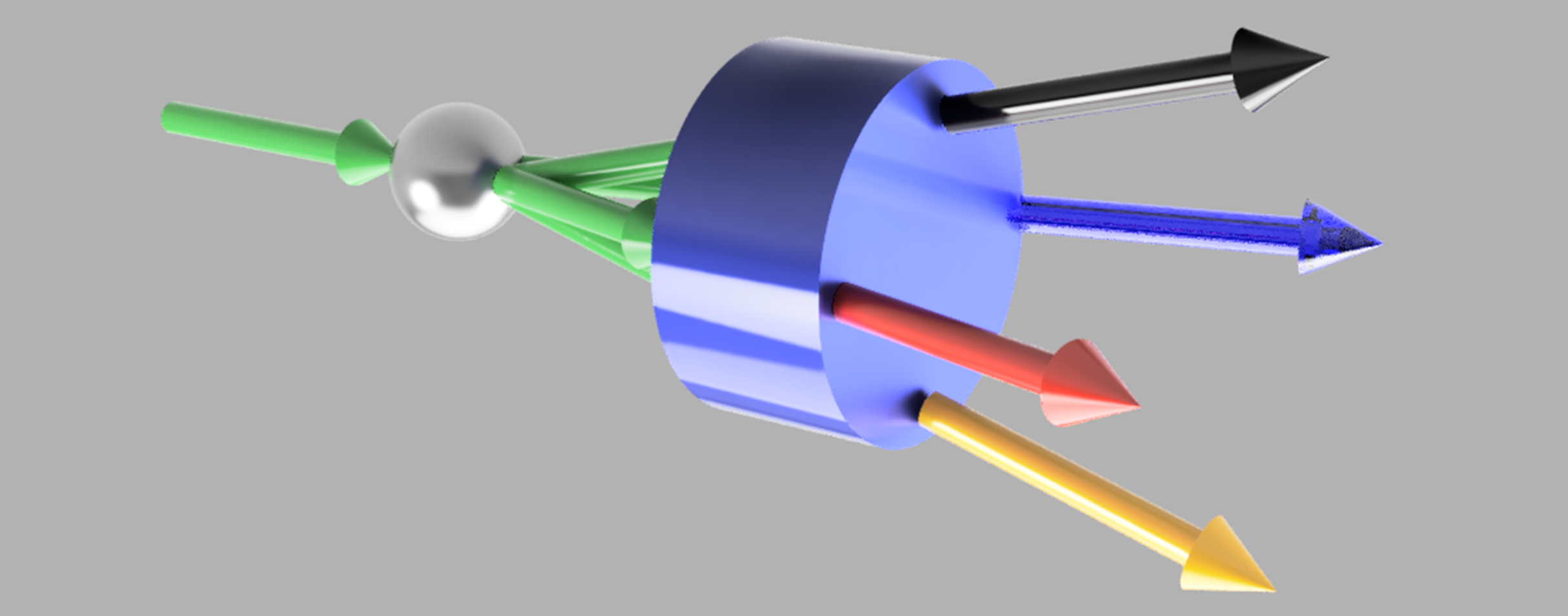}\caption{Schematic representation of a proposed setup to measure the longitudinal OAM generating mechanism in the weak interaction. An incident neutron (green arrow) with longitudinal polarization is scattered by a powder sample (metallic sphere) into a conical beam. This beam traverses through cylindrical $^{139}\mathrm{La}$ sample (blue), where the initially longitudinal spin couples to the transverse momentum component, thereby coupling the spin to the azimuthal coordinate. This spin-orbit coupling is visualized by the colors of the exiting beams, black and yellow represent spin right and spin left, while blue and red indicate spin up and spin down.}
	\label{OAMWeak}
\end{figure}
It is more illuminating to look at our Hamiltonian in the spectral domain, where the $x$ $y$ and $z$ coordinates have been Fourier transformed and converted to cylindrical coordinates ($k_r^2=k_x^2+k_y^2$ and $k_re^{i\phi}=k_x+ik_y$)
\begin{equation}\label{FourierHamiltonian}
H=k_r^2 +k_z^2- b'\begin{pmatrix}
&k_z \ \ &k_re^{-i\phi} \\ &k_re^{i\phi} \ \ &-k_z 
\end{pmatrix}
\end{equation}
or
\begin{equation}\label{FourierHamiltonian2}
H=k_r^2 +k_z^2- b'[k_r(\ell_+\sigma_-+\ell_-\sigma_+)+k_z\sigma_z]
\end{equation}
where $\ell_\pm = e^{\pm i\phi}$ and $\sigma_\pm$ represent the OAM and spin raising/lowering operators respectively. The eigenvalues of the Hamiltonian are given by $E_\pm=k_r^2+k_z^2\mp b'\sqrt{k_r^2+k_z^2}$, where the $\pm$ subscript denotes the up and down spin state respectively. Next we diagonalize the Hamiltonian with the transformation $H'=T^{-1}HT$, with $T=\begin{pmatrix}
&e^{-i\phi}\frac{[k_z+\sqrt{k_r^2+k_z^2}]}{k_r} \ \ &e^{-i\phi}\frac{[k_z-\sqrt{k_r^2+k_z^2}]}{k_r} \\
&1 \ \ &1
\end{pmatrix}$. In this transformed frame it is trivial to solve the time dependent Schroedinger equation: $H'\hat{\psi}'=\frac{\partial}{\partial t}\hat{\psi}'$:
\begin{equation}\label{SolutionDiag}
	\hat{\psi}'_\pm=\hat{a}_\pm e^{iE_\pm t}
\end{equation}
Finally we determine $\hat{\psi}$ by inverting the transform: $\hat{\psi}=T\hat{\psi}'$. The integration constants $\hat{a}_\pm$ are found using the initial conditions $\hat{\psi}_\pm(t=0)=\hat{c}_\pm(k_r,\phi,k_z)$. 
\begin{equation}\label{GeneralSolution}
\hat{\psi}=\begin{pmatrix}
&\hat{c}_+\cos(b'\sqrt{\epsilon}t)-i[\frac{\hat{c}_+k_z}{\sqrt{\epsilon}}-\frac{\hat{c}_-k_r}{\sqrt{\epsilon}}e^{-i\phi}]\sin(b'\sqrt{\epsilon}t) \\
&\hat{c}_-\cos(b'\sqrt{\epsilon}t)+i[\frac{\hat{c}_- k_z}{\sqrt{\epsilon}}-\frac{\hat{c}_+k_r}{\sqrt{\epsilon}}e^{i\phi}]\sin(b'\sqrt{\epsilon}t)
\end{pmatrix}
\end{equation}
with $\epsilon=k_r^2+k_z^2$ the kinetic energy parameter. We first note the more obvious but weaker mechanism by which orbital angular momentum is generated. This case occurs if the initial spin is prepared along the flight axis (z-direction i.e. $\hat{c}_-=0$), we find that over time the spin rotates until it is anti-parallel to the flight axis. During this process the wavefunction obtains one unit of $\hbar$ orbital angular momentum parallel to the propagation direction. This mechanism is proportional to $\frac{k_r}{\epsilon}$, which is equal to the beam divergence for small angles, hence this type of parity odd spin rotation is roughly a factor of 100 to 1000 weaker than previously observed parity odd spin rotation, where the spin is prepared perpendicular to the flight direction. Finally we wish to highlight the fact that this effect disappears under the (single) plane wave approximation where $k_r=0$, which is implicitly used in the usual neutron optics expressions. \par
The second mechanism, associated with the parity odd spin rotation which was observed in previous experiments~\cite{Forte1980,Heckel1984}, generates transverse orbital angular momentum. In this case the propagation direction is in the $k_r$ plane and the initial spin is perpendicular to the neutron momentum (for example $\hat{c}_-=0$). $k_r$ is now the dominating factor in the equation, while $\hat{c}_\pm\frac{k_z}{\sqrt{\epsilon}}\approx0$ (after integration ignoring the divergence). Under this approximation and $\sqrt{\epsilon}\approx k_r$ the expression \ref{GeneralSolution} becomes identical, within a phase factor of $i$, to the expression found for twisted neutrons in electric fields~\cite{Geerits2021}
\begin{equation}\label{SimplifiedSolution}
\hat{\psi}_\pm=\hat{c}_\pm\cos(b'k_rt)\pm i\hat{c}_\mp e^{\mp i\phi}\sin(b'k_rt)
\end{equation}
 We conclude that parity odd neutron spin rotations are compensated by an equal amount of orbital angular momentum generation to preserve total angular momentum. \par
When describing transverse orbital angular momentum it is helpful to expand $\hat{c}_\pm(k_r,\phi,k_z)$ as an infinite superposition of orbital modes. The amplitudes of these modes can be calculated via the azimuthal Fourier transform (AFT), which we have used in past work as well \cite{Geerits2021,Geerits2023,Geerits2024}
\begin{equation}\label{AzimuthalFourier}
	\hat{c}_\pm^\ell(k_r,k_z)=\frac{1}{2\pi}\int_0^{2\pi} d\phi \hat{c}_\pm(k_r,\phi,k_z) e^{-\mathrm{i}\ell\phi}
\end{equation} 
The raising and lower operators $e^{\pm i\phi}$ in equation \ref{SimplifiedSolution} simply shift the index of $\hat{c}^\ell_\pm$ by $\pm 1$ respectively. As a result the expectation value of $L_z$ is shifted by $\pm \hbar$, even in the case of transverse OAM. The amplitude or probability of the $\ell$th OAM mode, $A_\pm^{\ell}$, here termed the OAM distribution function, is given by
\begin{equation}\label{Amplitude}
	A_\pm^{\ell}=\int |\hat{c}_\pm^{\ell}|^2k_rdk_rdk_z
\end{equation}
which is the same in momentum space as in real space due to Parsevals theorem
\begin{equation}\label{Parseval}
	A_\pm^{\ell}=\int |\hat{c}_\pm^{\ell}|^2k_rdk_rdk_z=\int |c_\pm^{\ell}|^2rdrdz
\end{equation}
Hence we can conclude that if OAM is generated in momentum space it is also generated in real space, since the Fourier transform conserves this quantity. \par
\subsection{Production and Detection of Longitudinal OAM}
We have previously shown that longitudinal OAM production comes from the coupling between longitudinal spin and the transverse wavevector component $k_r$. Hence it is particularly effective to use a conical beam with a large opening angle to detect the longitudinal OAM effect. This can be achieved by passing a neutron beam through a powdered crystal. Diffraction from the powder causes the beam to be reflected on to multiple cones, where the opening angle of the cone corresponds to four times the Bragg angle of the respective lattice plane. After scattering the conical beam can be passed through a weak field, where the spin couples to the transverse momentum component. Since this momentum component is coupled to the angular coordinate on the cone, the spin after passing through the weak field will be aligned to the azimuthal coordinate, thereby producing a spin-orbit entangled state. This proposed setup is shown schematically in figure \ref{OAMWeak}.
\begin{figure}
	{\includegraphics[width=\textwidth/2]{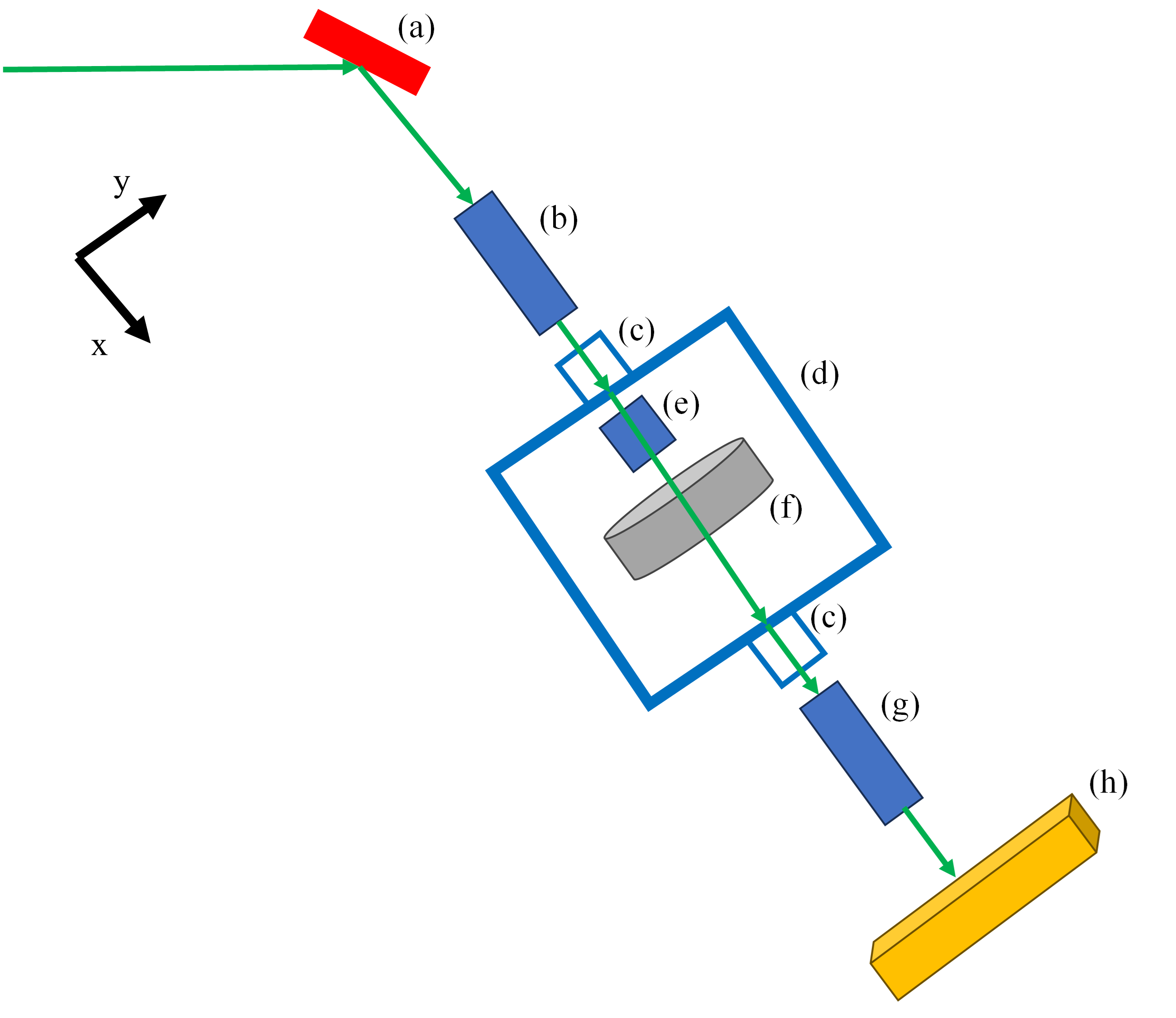}}\caption{Schematic representation of a top down view of the setup at IN3 used to demonstrate OAM generation and angular momentum conservation in the weak interaction. Thermal neutrons coming from the guide are Bragg diffracted and monochromatized ($\propto 1.8\mathrm{\AA}$) by a pyrolytic graphite crystal (a). After reflection neutrons are polarized by a bender polarizer (b). Next neutrons enter the zero field chamber (d) via the spin preparation/analysis coils (c). Inside the chamber a spin orbit state is prepared using a quadrupole (e), after which neutrons propagate through a $^{139}\mathrm{La}$ target. Behind the zero field chamber a bender (g) analyzes the neutron spin. Finally a position sensitive detector (h) resolves the spin-orbit pattern. The x and y axes are shown in black, while the z-axis points out of the page.}
	\label{LanthanumExperiment}
\end{figure}
Experimentally, such a setup would require a wide angle spin analyzer to determine the spin dependent azimuthal phase of the beam. In addition an appropriate powder sample is required to produce a beam with an appropriate opening angle and significant intensity. While these technical challenges can be overcome, for a demonstration experiment they can be circumvented by realizing that
the proposed experiment relies simply on the dot product between spin and momentum. In figure~\ref{OAMWeak} the momentum structure is conical and the spin longitudinal, however this can just as well be reversed, such that one uses a paraxial beam with a conical spin distribution across the beam. This leads to the exact same dot product $\vec{\sigma}\cdot\vec{p}$ as the case shown in the figure. Doing so removes the need for a wide angle spin analyzer as well as an appropriate powder sample. These requirements are replaced by the need for a spin optical device, which can produce the conical spin pattern. This can be accomplished using a magnetic quadrupole~\cite{Nsofini2016,Sarenac2018}. Note that this particular spin pattern corresponds to a spin orbit state. When passed through a weak field the spin-orbit pattern is rotated around the optical axis, despite this changing the spatial distribution of OAM and spin angular momentum, the total angular momentum of the neutron is conserved. Although this effect can be reproduced with a longitudinal magnetic field, the magnetic interaction does not generally conserve the total angular momentum of a neutron, since angular momentum can be dumped into the external magnetic field. 
\section{Methods}
\begin{figure}
	{\includegraphics[width=\textwidth/2]{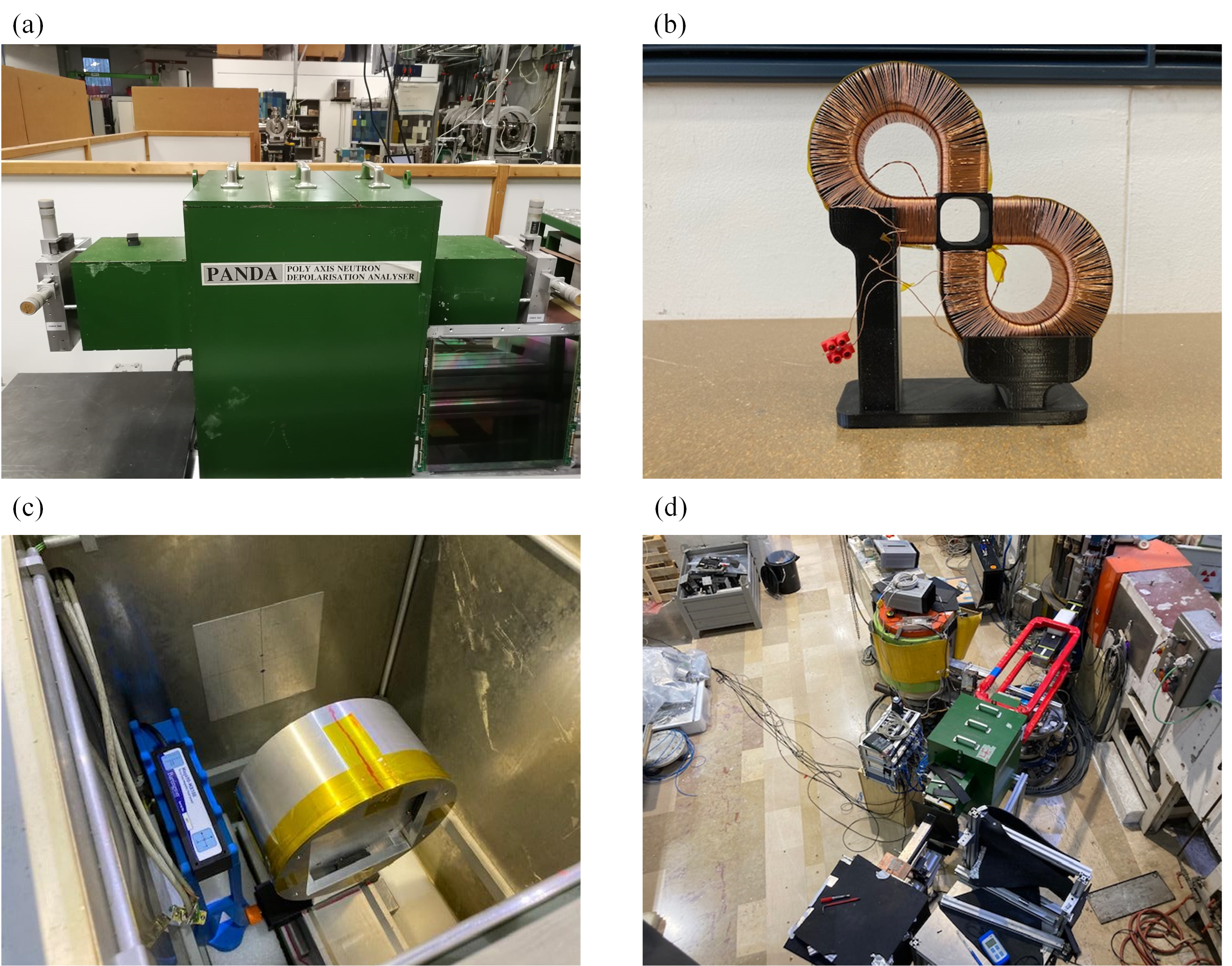}}\caption{Pictures of the zero field chamber PANDA (a), the magnetic quadrupole used to produce spin-orbit patterns (b), the $^{139}\mathrm{La}$ target inside of the zero field chamber (c) and the full setup at IN3 (d).}
	\label{LanthanumPictures}
\end{figure}
In this section we describe in more detail our experiment, demonstrating angular momentum conservation in the neutron nucleus weak interaction. Our experiment was conducted at the IN3 triple axis instrument of the Institute Laue Langevin~\cite{IN3DOI}.  Our setup, shown in figure~\ref{LanthanumExperiment}, uses a magnetic quadrupole to produce a spin-orbit pattern, as described in the previous section. This spin-orbit pattern is passed through a 5 cm long $^{139}\mathrm{La}$ sample, the strongest known amplifier of the neutron nucleus weak interaction.
\begin{figure}
	{\includegraphics[width=\textwidth/2]{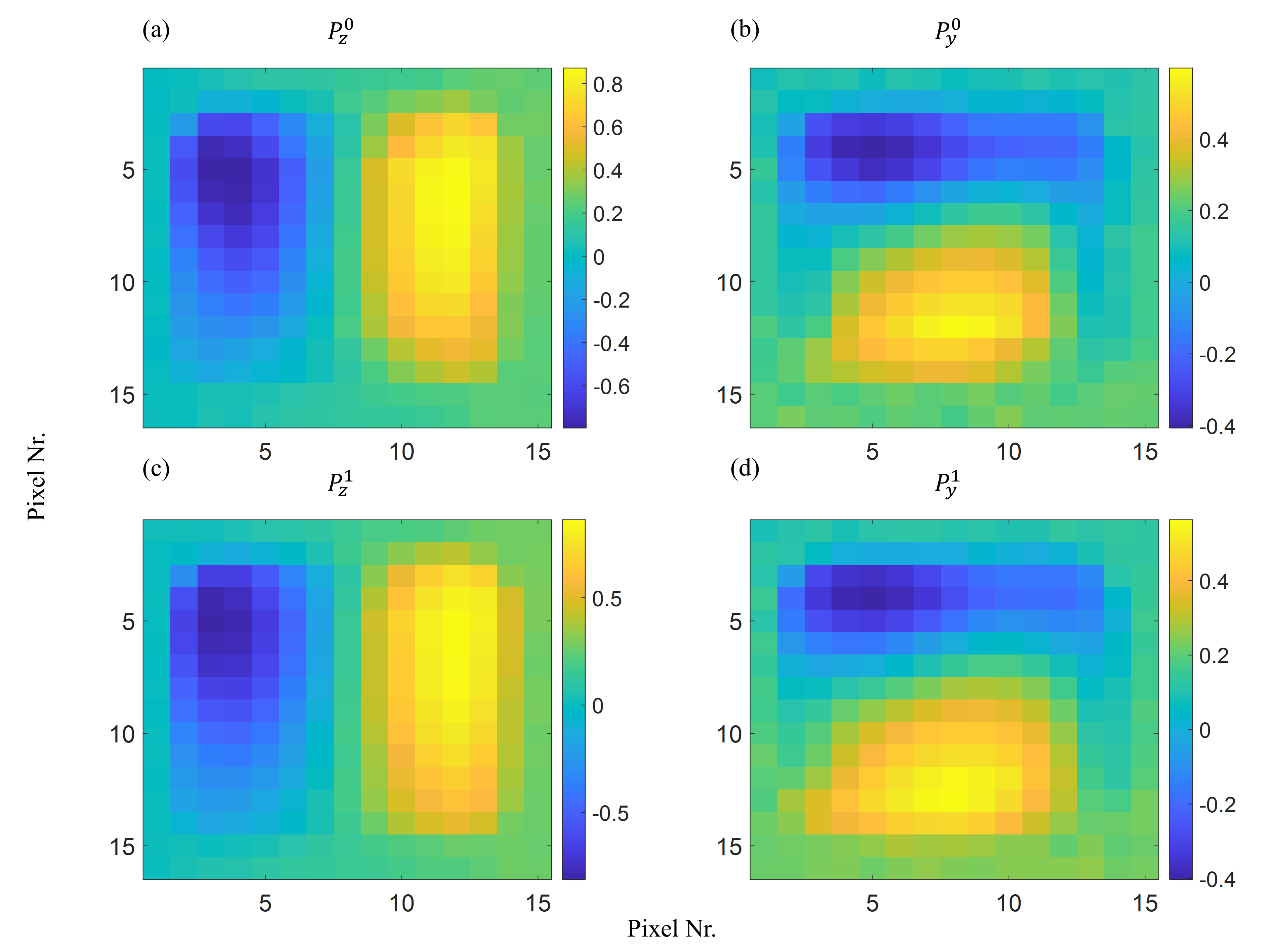}}\caption{Transverse polarization patterns produces using the quadrupole magnet. The z component is shown in (a), while the y component can be seen in (b). The same components are shown rotated due to the weak interaction in (c) and (d) respectively.}
	\label{LanthanumResults}
\end{figure}
Since the interaction strength is nonetheless weak ($2.2\cdot 10^{-4} \ \mathrm{rad\cdot cm^{-1}}$) at slow neuton energies far from the 0.7 eV resonance, both the $^{139}\mathrm{La}$ target and the quadrupole are contained within a zero field chamber. This zero field chamber, originally named Poly-Axis Neutron Depolarization Analyzer (PANDA)~\cite{bodnarchuk2008neutron}, provided by the Reactor Institute Delft, consists of three shielding layers, two passive metal (steel and $\mu$ metal) and one internal active stage. This active shielding is realized by three pairs of orthogonal Helmholtz coils. The coils are driven using a feedback loop which is controlled using a fluxgate inside the zero field chamber. When activated, the residual field in the chamber can be suppressed to $\approx$100 nT. The $^{139}\mathrm{La}$ target sits in an aluminum and metal housing $\mu$, to further suppress the field and prevent oxidation. In addition to the magnetic shields PANDA has a pair of mezei and v-coils on the entrance and exit of the chamber, so that the neutron spin can be prepared and analyzed along any direction. The instrument uses a pyrolytic graphite monochromator in a thermal guide to reflect neutrons from the thermal peak into the setup. Neutrons are then polarized via a bender and enter into the PANDA chamber. While entering the chamber the neutron spin is rotated in the longitudinal (x) direction. Inside the chamber a quadrupole produces the spin orbit pattern. To resolve the spin orbit pattern both transverse spin components
need to be analyzed using the exit coils and the bender and the resulting intensity
pattern must be imaged using a position sensitive detector. In our case the position
sensitive detector consists of a Helium-3 wire chamber with 2x2 $\mathrm{mm^2}$ resolution~\cite{clergeau2016information}. At thermal energies the efficiency is around 80\%. When the $^{139}\mathrm{La}$ inserted in the beam the spin orbit pattern is rotated. By subtracting the spin pattern measured without $^{139}\mathrm{La}$ from the pattern measured with $^{139}\mathrm{La}$ one can obtain the change in neutron spin, as well as any possible changes in beam OAM. This allows us to determine whether or not angular momentum is conserved within the neutron beam. The magnetic quadrupole is realized by a figure eight coil. Pictures of this quadrupole, the setup, PANDA and the packed $^{139}\mathrm{La}$ target are shown in figure \ref{LanthanumPictures}.

\section{Results}
\begin{figure}
	{\includegraphics[width=\textwidth/2]{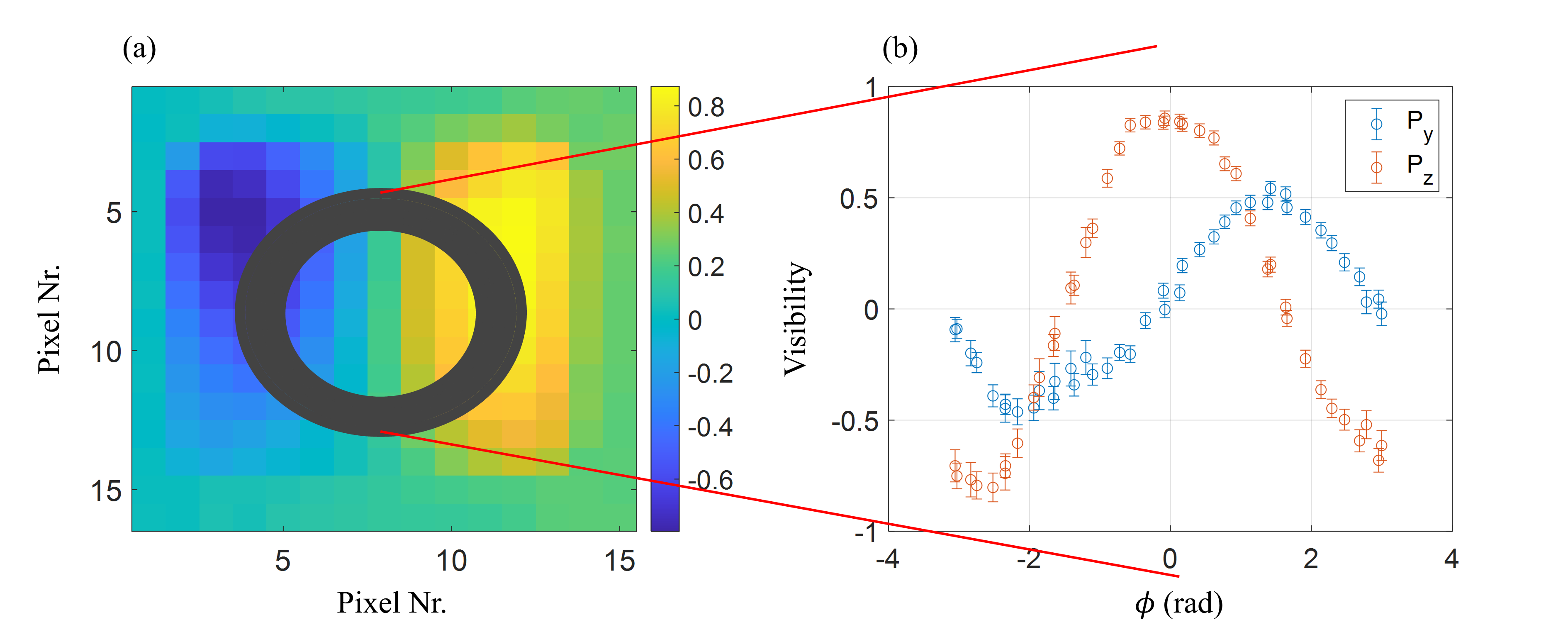}}\caption{Polarization against azimuthal angle (b) for a certain circular cut out of the two dimensional polarization pattern (a). The circular cut taken is shown as a grey annulus. The pixels that lay inside of the ring are used to construct the polarization plots seen in (b).}
	\label{LaCircles}
\end{figure}

\begin{figure}
	{\includegraphics[width=1\textwidth/2]{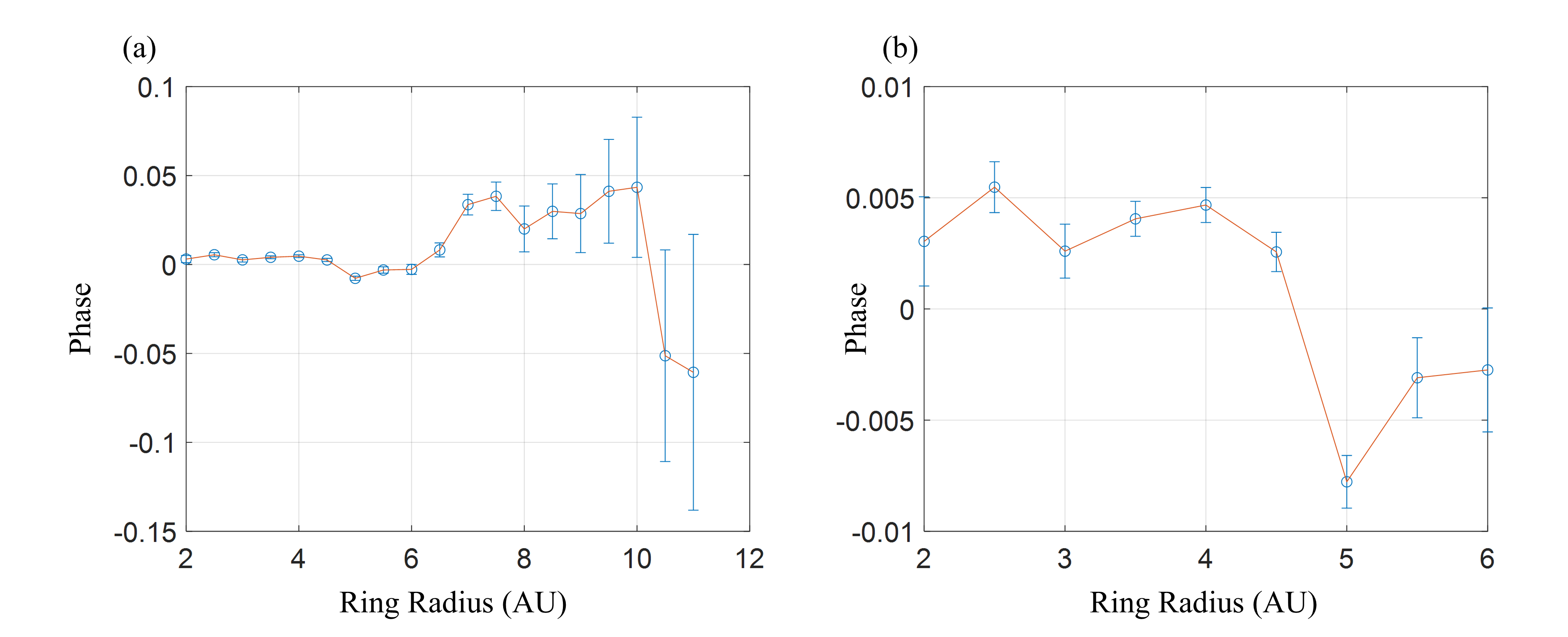}}\caption{Image rotation phase plotted against the ring radius. Inset (a) shows the entire dataset, while inset (b) shows a zoomed variant.}
	\label{LanthanumPhases}
\end{figure}

\begin{figure}
	{\includegraphics[width=1\textwidth/2]{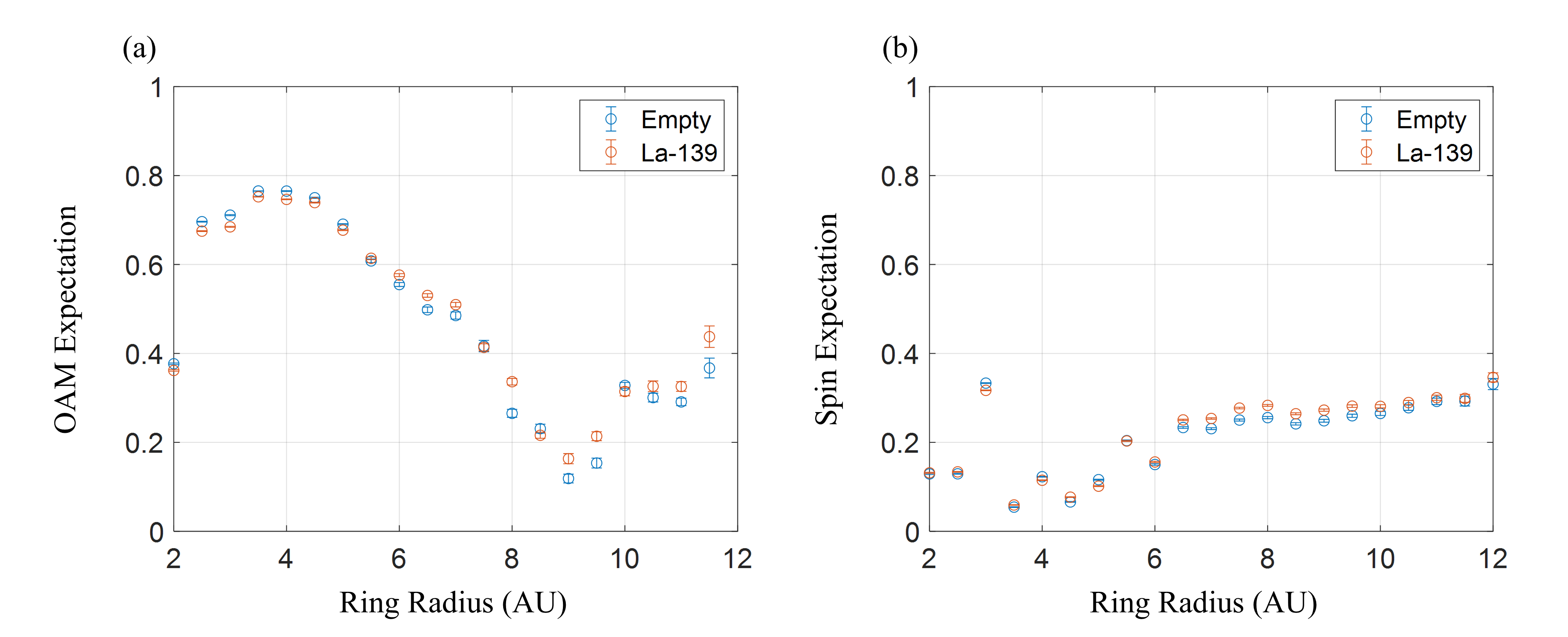}}\caption{(a) Expectation value of the orbital angular momentum operator as function of beam radius for the empty beam (blue) and the case where $^{139}\mathrm{La}$ is inserted into the beam (red) and (b) the expectation value of the spin angular momentum as a function of beam radius.}
	\label{AMConservation}
\end{figure}

The transverse polarization patterns produced using the quadrupole are shown in the top insets of figure~\ref{LanthanumResults}. We note that this is the first neutron spin-optical experiment demonstrating twisted spin patterns using a magnetic quadrupole. These spin orbit patterns rotated by the weak interaction are shown in the bottom insets. This rotation cannot be seen upon visual inspection. Hence additional analysis is required, to determine the rotation angle. 
To this end we make circular cuts out of the y and z polarizations and perform AFTs on each cut. For illustration purposes we have plotted the polarization against the azimuthal angle $\phi$ for one such circular cut in figure \ref{LaCircles}. 
Here the y and z polarization are shown for a cut out of the spin-orbit pattern produced by the quadrupole without $^{139}\mathrm{La}$ in the beam. After performing these cuts for various different radii, we apply the azimuthal Fourier transform for $\ell=1$ to extract the phase of this particular beam OAM mode from $P_y +\mathrm{i}Pz$. Finally the extracted phases are averaged over the different radii and the average phase determined without the $^{139}\mathrm{La}$ sample is subtracted from the phase determined with the $^{139}\mathrm{La}$ sample in the beam. The phase is shown as a function of the ring radius in figure \ref{LanthanumPhases}. When calculating the weighted average over these phases we find an average image rotation of $2.4\pm 0.27\cdot 10^{-3}$ radians.
This is about two times the rotation expected due to the weak interaction, hence it
is reasonable to assume that the $^{139}\mathrm{La}$ sample includes a magnetic component
arising from ferromagnetic impurities in the sample. This can be isolated by rotating
the $^{139}\mathrm{La}$ sample by 180° such that the surface normal is inverted. This inverts
the effective magnetic field due to impurities seen by the neutrons, while leaving the weak contribution unchanged. Doing so we find that the magnetic contribution to the spin rotation was $1.1\cdot 10^{-3}$ radians. Hence we can conclude that the contribution coming from the weak interaction is within the expected range for $^{139}\mathrm{La}$. We find, as expected, that our value for thermal neutrons does not differ from that found in cold neutrons \cite{Heckel1984}. In the previous section, we claimed to be able demonstrate that the angular momentum of the neutron beam is conserved in the weak interaction. In figure \ref{AMConservation} we show the orbital and spin angular momentum of the beam with and without the $^{139}\mathrm{La}$ sample in the beam as a function of beam radius. The beam OAM is calculated simply via the expectation value of $-\mathrm{i}\frac{\partial}{\partial \phi}$, while the spin expectation value is approximated by the average magnitude of the measured 2D polarization vector. The difference between the total angular momentum of the empty beam and the beam with $^{139}\mathrm{La}$ in it is zero to within error ($\pm 4 \cdot 10^{-4}$).
\section{Conclusion}
In our theoretical framework we have demonstrated that the neutron nucleus weak interaction in a very similar way to the Schwinger interaction generates OAM in neutrons by spin to OAM conversion and total angular momentum conservation. Furthermore using thermal neutrons we have experimentally confirmed previous experiments done with cold neutrons on $^{139}\mathrm{La}$ that the weak interaction in this sample, amplified by a p-wave resonance, induces a parity non conserved spin rotation of $(2.2\pm 0.5)\cdot 10^{-4} \ \mathrm{rad}\cdot\mathrm{cm}^{-1}$. By coupling spin and beam OAM this spin rotation can be manifested as an image rotation which we have measured experimentally. Our experiment simulates a neutron beam with a conical momentum distribution by introducing a conical spin distribution using a magnetic quadrupole. Finally we showed that in this experiment total angular momentum is conserved in the neutron to within statistical error.
\section{Acknowledgments} N. G. and S. S. acknowledge funding from the Austrian
Science Fund (FWF), grant DOI: 10.55776/P34239, in addition
N. G., K. S. and W. M. S. recognize support by the US Department of Energy (DOE) grant DE-SC0023695. All authors are grateful for the beam time made available by the ILL~\cite{IN3DOI}
\bibliography{SpinOrbitBibliography}
\end{document}